hep-ph/0511015 

\documentclass[11pt]{cernrep}
\usepackage{graphicx}
\begin{document}
\title{Quasi-particle perspective on QCD matter and critical end point effects}
\author{Marcus Bluhm$^1$\, and Burkhard K\"ampfer$^{1,2}$}
\institute{$^1$ Institut f\"ur Kern- und Hadronenphysik, 
  Forschungszentrum Rossendorf, PF 510119, 01314 Dresden, \\ \hskip 2.8mm Germany\\
$^2$ Institut f\"ur Theoretische Physik, TU Dresden, 01062 Dresden, Germany}

\maketitle
\begin{abstract}
  Our quasi-particle model is compared with recent lattice QCD data at finite 
  temperature and baryon number density with emphasis on the coefficients 
  in the Taylor series expansion of thermodynamic observables. The inclusion 
  of static critical end point effects into the equation of state is discussed.
\end{abstract}
\section{Introduction}
\label{sec:Intro}

The QCD phase diagram exhibits an astonishingly rich phase structure. The (pseudo-) 
critical line, which separates the phase dominated by quark and gluon degrees of freedom at 
large temperature $T$ and baryo-chemical potential $\mu_B$ from the one dominated by hadrons 
and resonances, has been investigated by means of lattice gauge 
theory~\cite{All02,deForc02,deForc03,Fodor}. From theoretical 
reasoning~\cite{Hala98,Steph} for finite quark masses, the phase transition is of first order at 
finite $T$ and large $\mu_B$ ending in a critical point of second order. For smaller $\mu_B$, 
thermodynamic observables change rapidly but continuously indicating a crossover regime. There, the 
equation of state (EoS) has been computed for $N_f=2$ quark flavours~\cite{Karsch00,All05}. 
While the location of the critical end point (CEP) showing a strong quark mass 
dependence~\cite{deForc03,Schmidt03} was determined by first principle QCD 
evaluations~\cite{Fodor,Gavai05}, the extension of the critical region is fairly unknown. 
Lattice QCD studies of the volume dependence of the Binder cumulant~\cite{deForc03,Karsch01} 
indicate that the CEP belongs to the static universality class of the 3D Ising model. At present, 
many investigations aim to study implications of such a fundamental issue of QCD. In 
particular, observable consequences of the occurrence of the CEP as novel feature of QCD 
are discussed~\cite{Asakawa00}. Heaving in mind the successful hydrodynamical description of 
the expansion stage in heavy-ion collisions, one intriguing problem concerns the manner the 
EoS of strongly interacting matter becomes modified by the CEP. This is the subject of the present 
contribution.

In~\ref{sec:Taylorcoeffs}, the quasi-particle model is shortly reviewed and compared with recent 
lattice QCD results. We focus on the Taylor series expansion coefficients~\cite{Karsch00,All05}. 
In~\ref{sec:CEP}, CEP effects on the EoS are discussed for a toy model and for our  
QCD based quasi-particle model. The results are summarized in~\ref{sec:conclusions}.

\section{Taylor series expansion of the EoS}
\label{sec:Taylorcoeffs}

Thermodynamic observables can be expressed as Taylor series expansions in powers of $\mu_B /T$. 
Accordingly, the pressure is decomposed into
\begin{equation}
  \label{equ:pres1}
  p(T,\mu_B)=T^4\sum_{n=0}^\infty
  c_{2n}(T)\left(\frac{\mu_B}{3T}\right)^{2n}, 
\end{equation}
where $c_0(T)=p(T,\mu=0)/T^4$ and 
$c_k(T)=\left.\partial^k p/\partial\mu^k\right|_{\mu=0}T^{k-4}/k!$ with $\mu=\mu_B /3$. 
In~\cite{Karsch00,All05}, the Taylor expansion coefficients up to $c_6(T)$ have been presented basing on 
first principle QCD evaluations. 

Achieving a flexible parametrization of the EoS, we formulated a model which
describes the quark-gluon fluid in terms of quasi-particle excitations~\cite{Pesall,Blu05}
\begin{equation}
  \label{equ:pres2}
  p(T,\mu)=\sum_{i\,=\,q,g} p_i(T,\mu) - B(T,\mu) .
\end{equation}
Here, $p_i$ denote thermodynamic standard expressions for quarks and transverse gluons with 
dynamically generated self-energies $\Pi_i$ and a non-perturbative effective coupling $G^2(T,\mu)$ 
as essential input. Thermodynamic self-consistency is ensured through the stationarity 
conditions $\delta p/\delta \Pi_i=0$, imposing in turn 
conditions onto $B(T,\mu)$. From Maxwell's 
relation for $p$, a flow equation for $G^2(T,\mu)$ follows~\cite{Pesall} 
\begin{equation}
  \label{equ:flow}
  a_\mu\frac{\partial G^2}{\partial\mu} + a_T\frac{\partial
  G^2}{\partial T} = b .
\end{equation}
Knowing $G^2$ on an arbitrary curve in the $T$ - $\mu$ plane,~(\ref{equ:flow}) 
can be solved as a Cauchy 
problem. For convenience, we adjust $G^2(T,\mu=0)$ to lattice data at $\mu=0$ enabling 
a mapping into the finite chemical potential region via~(\ref{equ:flow}). 
The Taylor expansion coefficients follow straightforwardly from~(\ref{equ:pres2}) as 
integral expressions involving $G^2$ and higher order derivatives of the effective coupling 
at vanishing chemical potential. The latter can be computed by exploiting the flow 
equation~(\ref{equ:flow}) (cf.~\cite{Blu05} for details). In Fig.~1, 
a fairly good agreement between Taylor expansion coefficients from the quasi-particle 
model and lattice results is shown.
\begin{figure}[t]
  \vskip -.5cm
  \label{fig:coeffs}
  \begin{minipage}[t]{12cm} 
    \begin{minipage}[h]{3.5cm}
      \includegraphics[scale=0.26,angle=-90.]{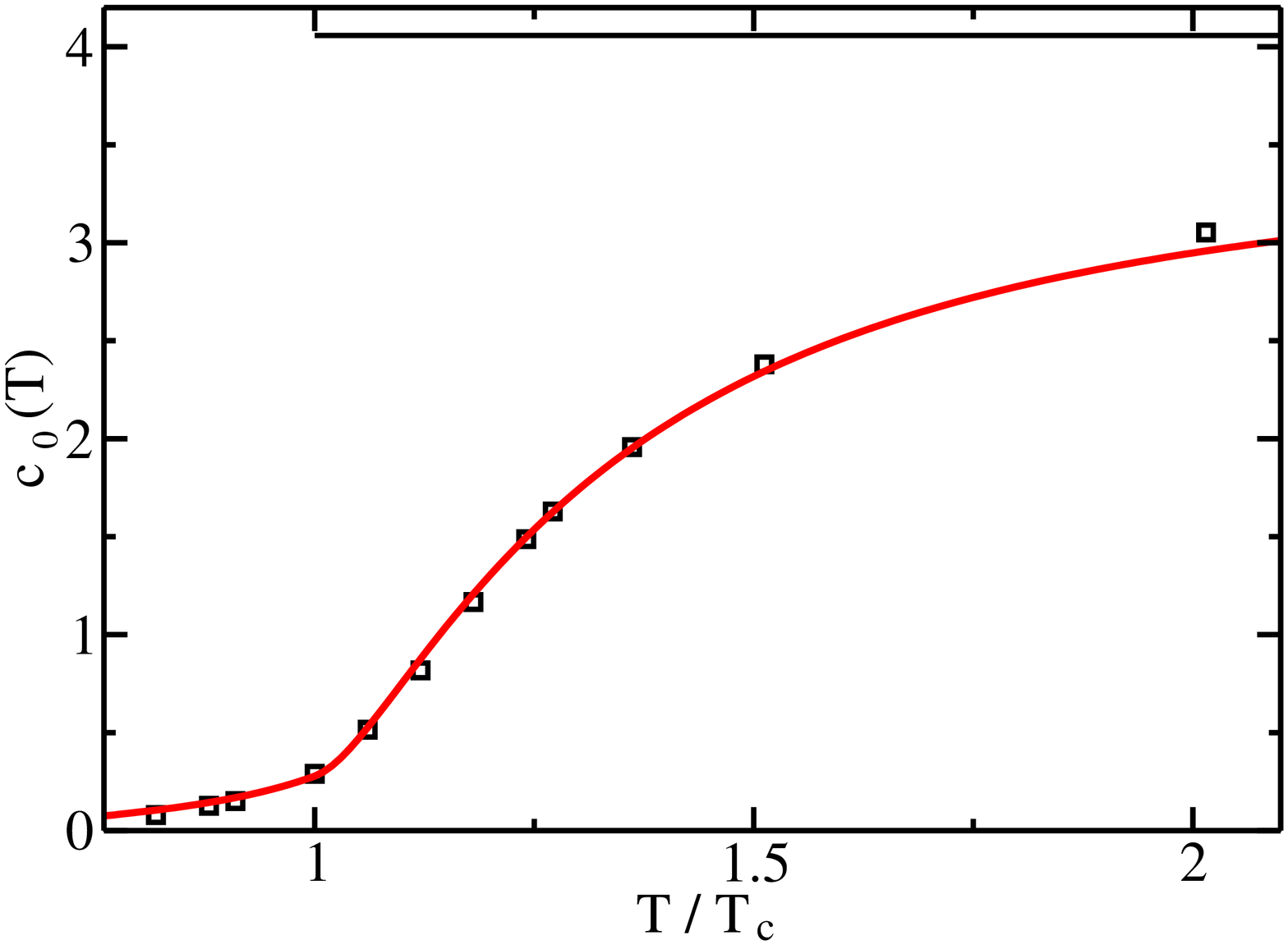}
    \end{minipage} 
    \hskip 4.9cm
    \begin{minipage}[h]{3.5cm}
      \includegraphics[width=1.63\linewidth,angle=-90.]{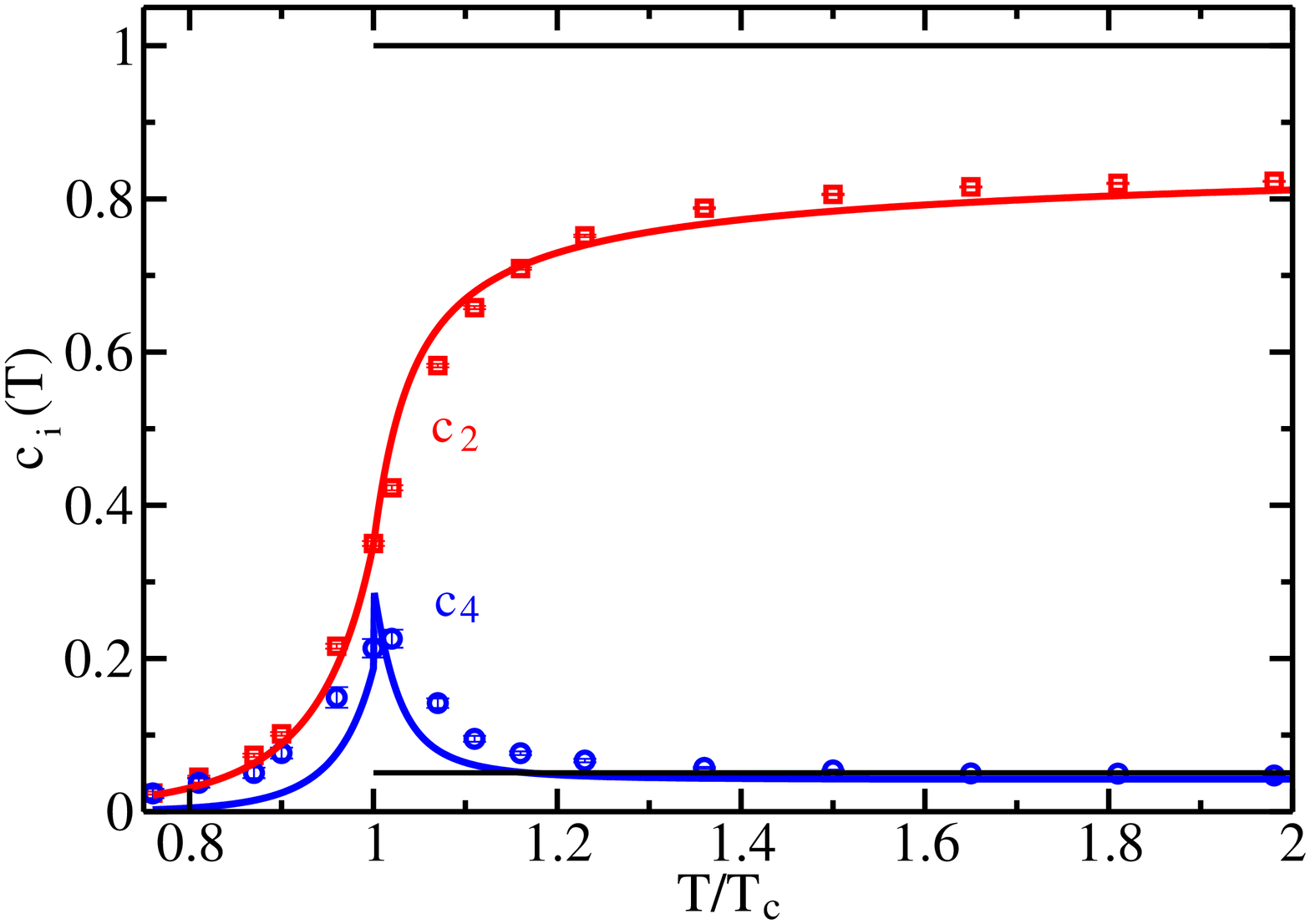}
    \end{minipage} \\
    \begin{minipage}[t]{3.5cm}
    \vskip -4mm
      \includegraphics[width=1.74\linewidth,angle=-90.]{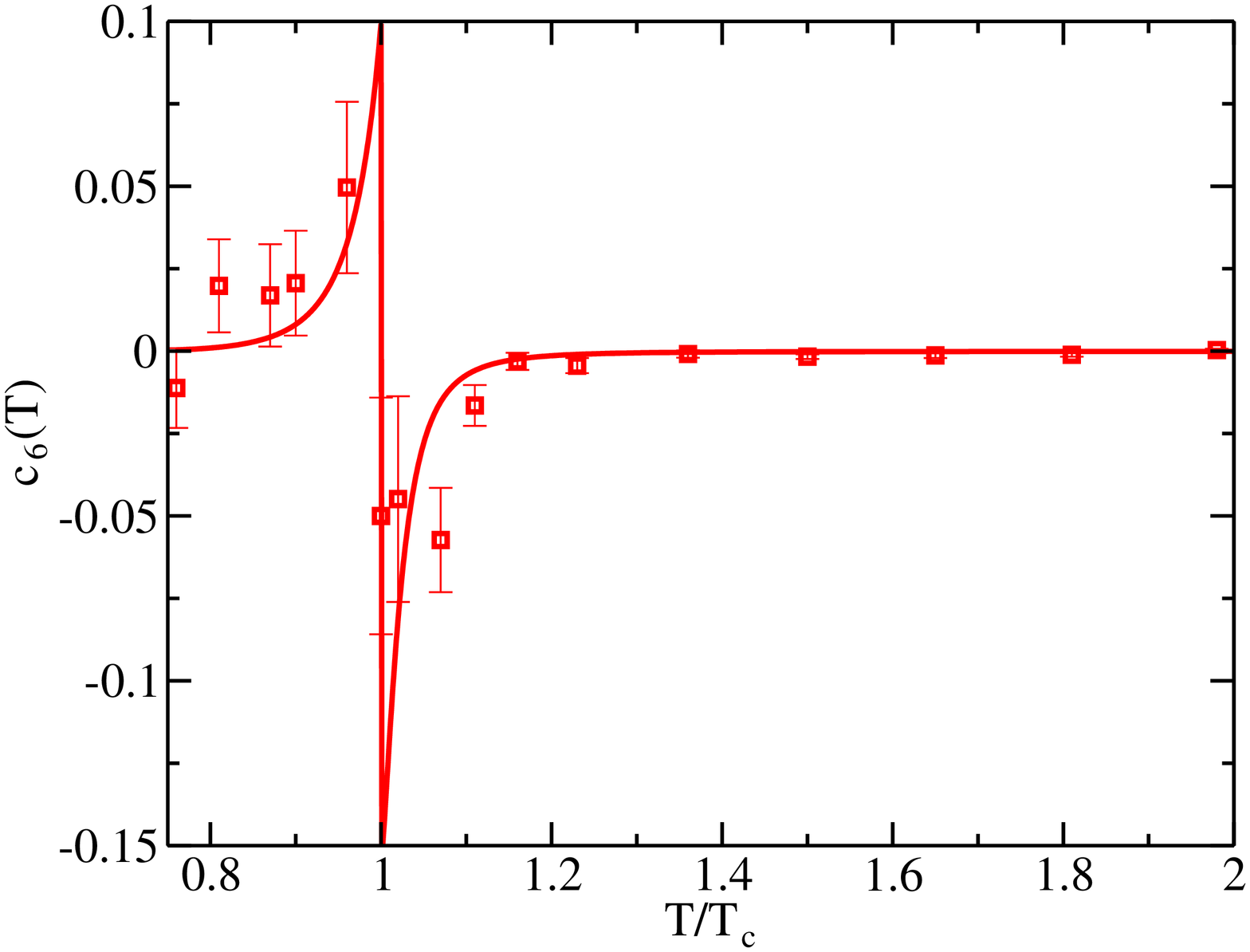}
    \end{minipage}
    \hskip 5.5cm
    \begin{minipage}[t]{6cm}
      \vskip 1.3cm
      \caption{Taylor expansion coefficients $c_{0,2,4}$ (top row, horizontal lines indicate 
        corresponding coefficients for a non-interacting gas above $T_c$ as previously often used to describe 
        the quark-gluon plasma) and $c_6$ (left panel in bottom 
        row) of~(\ref{equ:pres1}) as a function of $T/T_c$. Lattice QCD data (symbols) 
        from~\cite{Karsch00,All05}. 
      }
    \end{minipage}
  \end{minipage}
\end{figure}
In particular, the peak in $c_4(T)$ and the dipole structure in $c_6(T)$ are reproduced. Adjusting 
the parametrization of the effective coupling onto $c_0(T)$ dictates a change in $G^2(T,\mu=0)$ 
at $T_c=170$ MeV from a regularized logarithmic dependence (resembling the perturbative behaviour 
at large $T$) into a linear dependence. This change in the curvature of $G^2$ can be 
considered as implemented phase transition and is responsible for the pronounced structures 
in $c_{4,6}(T)$. In fact, these structures disappear when neglecting higher order derivatives of 
the effective coupling in the integral expressions of $c_{4,6}(T)$ serving for a test 
of the flow equation~(\ref{equ:flow}). 

\section{Critical end point}
\label{sec:CEP}

Starting from a thermodynamic potential, e.~g. Gibbs free enthalpy, it can be decomposed 
into an analytic and a non-analytic part where the latter is related to phase 
transitions and critical phenomena~\cite{Gebhardt80}. Accordingly, the EoS formulated in terms 
of the entropy density is given through $s = s_a + s_n$.
Here, the analytic contribution $s_a$ has to be adjusted to the known EoS outside of the 
critical region. The non-analytic part $s_n$ should embody the feature of being continuous left to the 
CEP (i.~e. at small chemical potentials) whereas on the right (at large chemical potentials) it 
generates a first order phase transition. A convenient parametrization of $s_n$ for the 3D Ising model 
characterized by a set of critical exponents is given in~\cite{Guida97}. 
Still, the corresponding variables employed usually in 
condensed matter physics including the order parameter need to be mapped into the $T$ - 
$\mu_B$ plane in the vicinity of the CEP. Details of this mapping and a useful formulation 
of the entropy density contribution can be found in the pioneering work~\cite{Nonaka05} 
we rely on. In the following, we estimate the phase border line to be given by 
$T_c(\mu_B)=T_c\left(1+\frac{1}{2}d(\mu_B / 3T_c)^2\right)$ with $d=-0.122$ 
according to~\cite{All02,deForc03} and locate the CEP at $\mu_{B,c}=360$ MeV in agreement 
with~\cite{Fodor}. 
\begin{figure}[t]
  \vskip -.5cm
  \label{fig:transBag}
  \includegraphics[scale=0.29,angle=-90.]{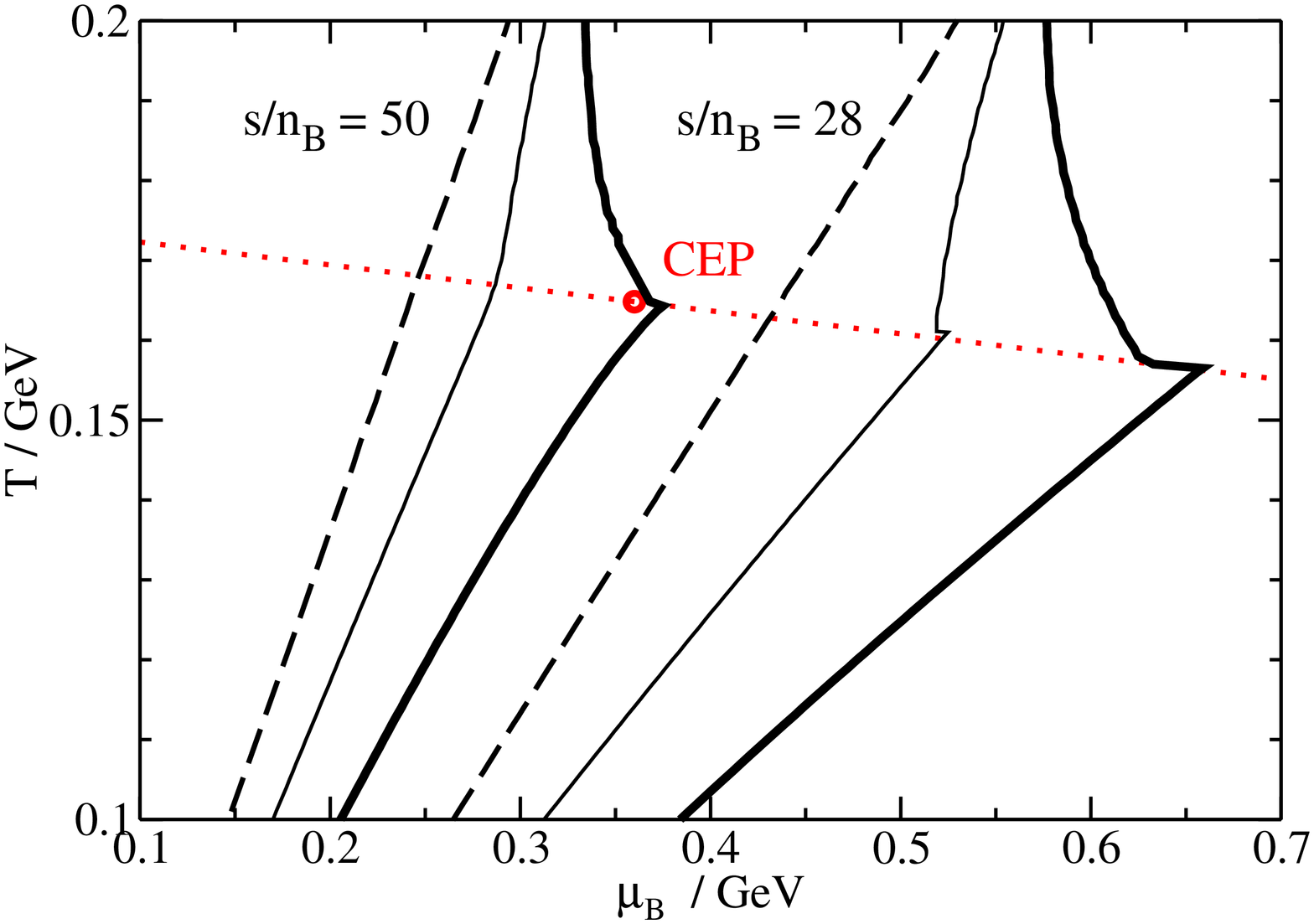}
  \hskip -2mm
  \includegraphics[scale=0.29,angle=-90.]{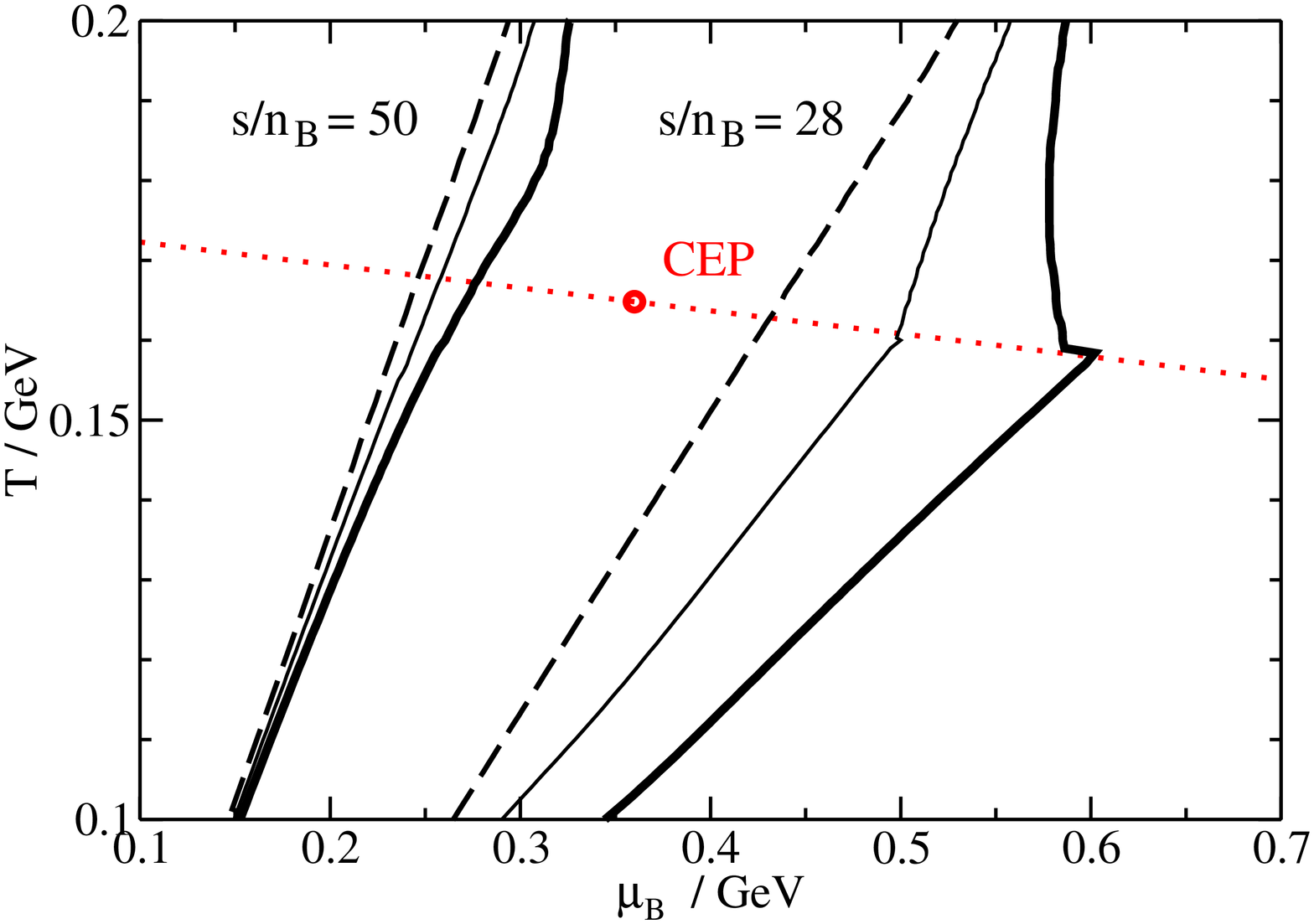}
  \caption{Isentropic trajectories of the toy model~(\ref{equ:entrdens}) depending on 
    the strength parameter $A$ for $s/n_B=50,\, 28$. Dashed, thin and solid lines exhibit 
    results for $A = 0.,\, 0.5,\, 1.0\,$ respectively. Dotted lines represent the tangent on the 
    estimated (pseudo-) critical line at the CEP. 
    Left panel: $\Delta T = 100$ MeV, $\Delta\mu_B = 200$ MeV, $D=0.15\,$ ; $\,$
    right panel: $\Delta T = 10$ MeV, $\Delta\mu_B = 10$ MeV, $D=0.06$.}
\end{figure}

As a simple toy model, let us employ the first terms in~(\ref{equ:pres1}), however, with 
constant expansion coefficients. 
The entropy density contributions are given by 
\begin{equation}
  \label{equ:entrdens}
  s_a(T,\mu_B)=4\bar{c}_0T^3+\frac{2}{9}\bar{c}_2\mu_B^2 T\,,\quad 
  s_n(T,\mu_B)=\frac{2}{9}\bar{c}_2\mu_B^2 T\, A \,\tanh \left(S_c(T,\mu_B)\right)
\end{equation}
with $\bar{c}_0=(32+21N_f)\pi^2 /180$, $\bar{c}_2=N_f /2$ and $N_f=2$. $n_B$ follows 
from~(\ref{equ:entrdens}) via standard thermodynamic relations (cf.~\cite{Nonaka05}) 
with integration constant $n_B(0,\mu_B)=\frac{4}{3}\bar{c}_4(\mu_B/3)^3$ where 
$\bar{c}_4=N_f / 4\pi^2$. The ansatz for $s_n$ has been chosen such that $s_n\rightarrow 0$ for 
$T\rightarrow 0$ and the net baryon density vanishes at $\mu_B=0$. The 
parameter $A$ describes the strength of the non-analytic contribution 
in the EoS. We apply the same $S_c(T,\mu_B)$ as in~\cite{Nonaka05} assuming a fairly large 
critical region parametrized by $\Delta T = 100$ MeV, $\Delta\mu_B = 200$ MeV and a 
stretch factor $D = 0.15$. Hence, CEP effects on the EoS and in particular on 
isentropic trajectories $s/n_B=const$ in the $T$ - $\mu_B$ plane can be demonstrated. 
In Fig.~2, 
the influence of the strength parameter $A$ on the behaviour of isentropic trajectories is exhibited. 
For increasing $A>0$, the trajectories for large $s/n_B$ tend to be attracted towards larger $\mu_B$ 
due to the presence of the CEP. In fact, the CEP acts as an attractor on trajectories on the left 
whereas on the right a repulsive impact is found. Evidently, the curves on the right side 
of the CEP display the existence of the first order phase transition. 
By shrinking the critical region to $\Delta T = 10$ MeV, $\Delta\mu_B = 10$ MeV and $D=0.06$, 
the influence of the CEP decreases (right panel of Fig.~2) 
in comparison with 
the results obtained for a large critical region (left panel of Fig.~2). 
In particular, the sections on the hadronic side become less affected when decreasing the extension 
of the critical region. 

The parameters in the non-analytic entropy density contribution and 
in particular $A$ have to be chosen such that standard thermodynamic consistency 
conditions are satisfied~\cite{Nonaka05}. 
Accordingly, during the adiabatic expansion of the system and its 
related cooling, both $n_B$ and $s$ must decrease. For $A<0$ with trajectories for large $s/n_B$ bent to 
smaller $\mu_B$ due to the CEP inclusion, however, these conditions are 
violated in the vicinity of the first order transition line. 
Therefore, the pattern of the isentropic trajectories as exhibited in Fig.~2 
\begin{figure}[t]
  \label{fig:isentrops}
  \includegraphics[scale=0.29,angle=-90.]{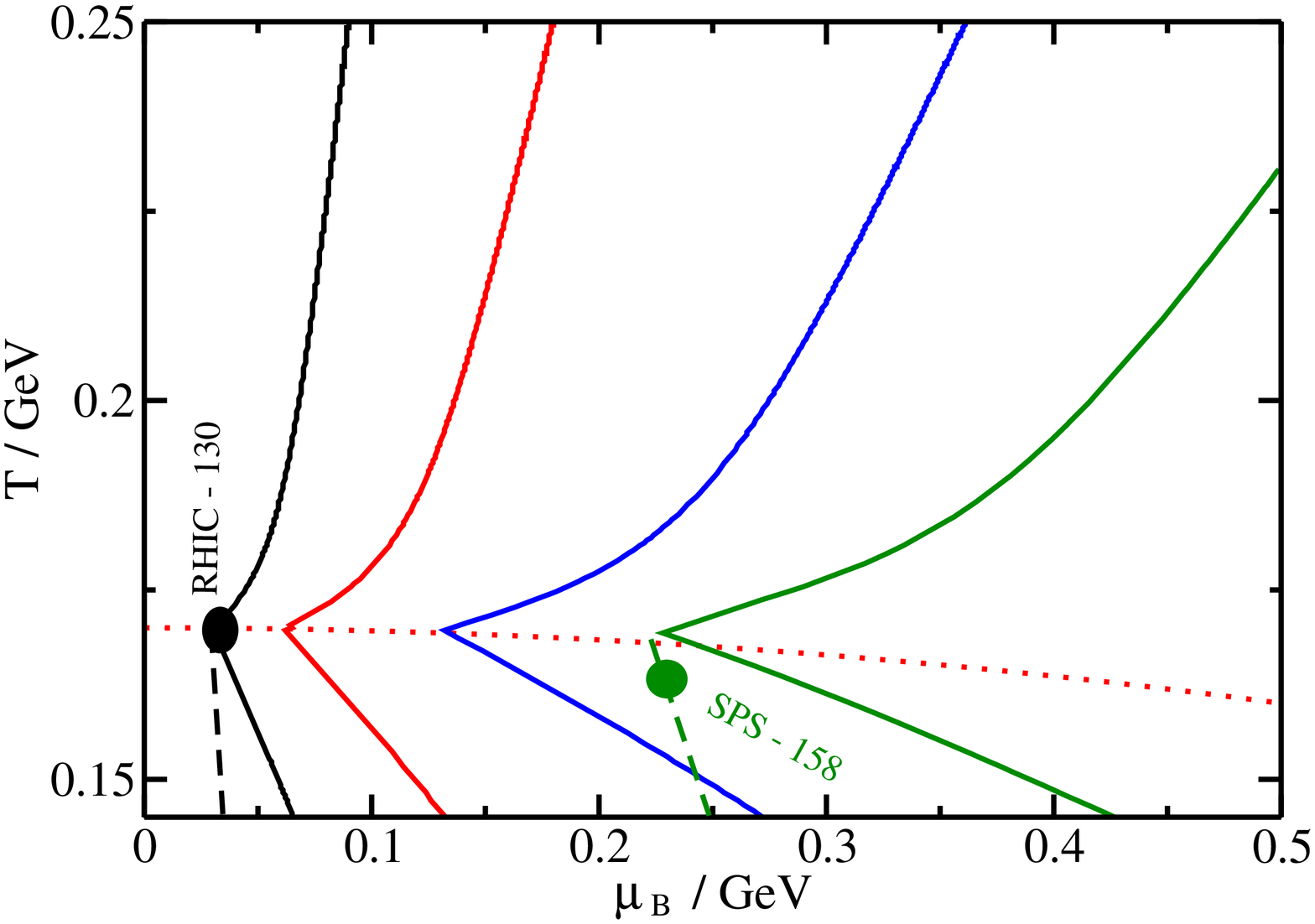}
  \hskip -2mm
  \includegraphics[scale=0.29,angle=-90.]{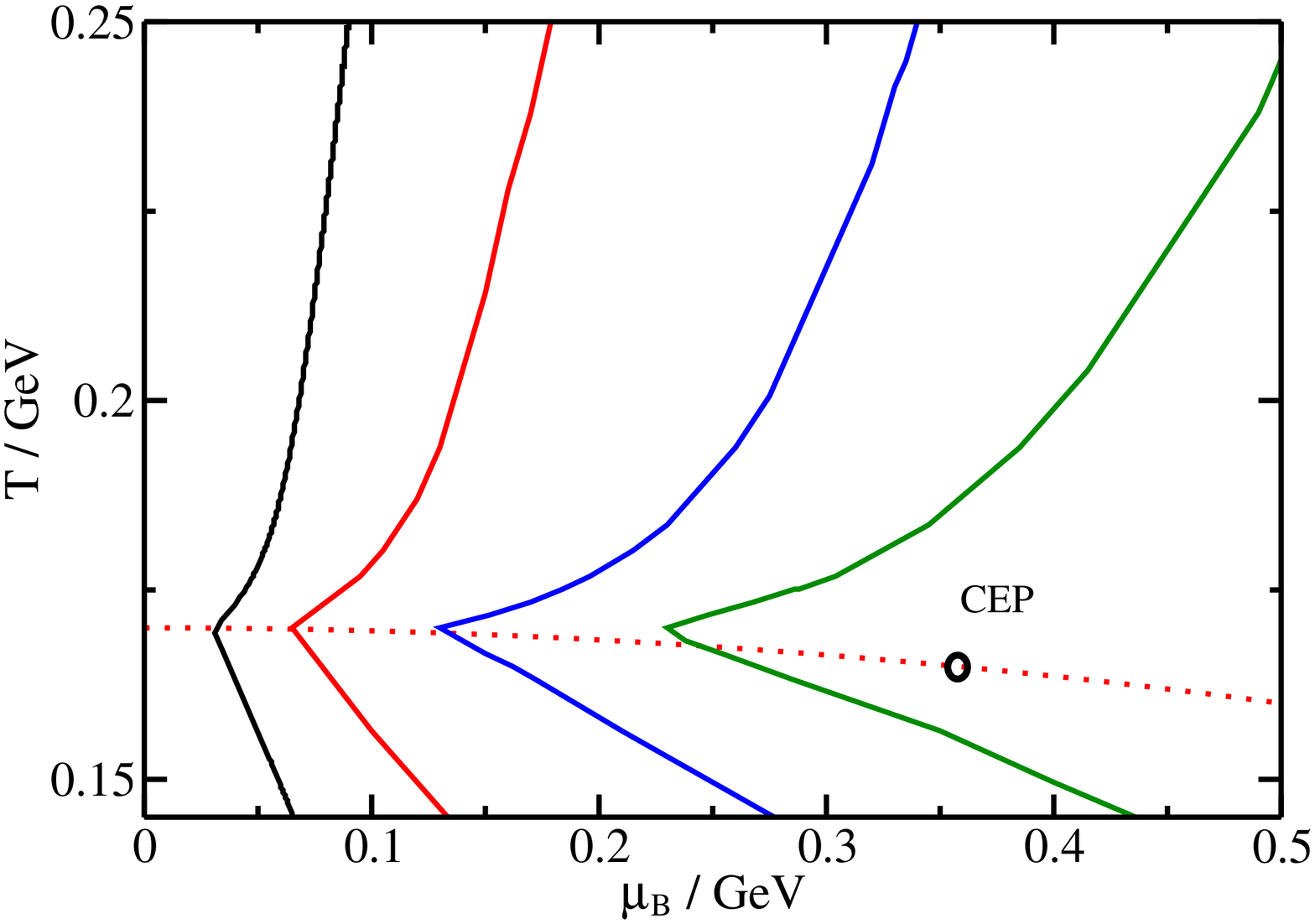}
  \caption{Isentropic trajectories in the truncated quasi-particle model adjusted to lattice 
    QCD~\cite{Karsch00,All05} without CEP (left panel) and with CEP effects included (right 
    panel) parametrized by $\Delta T = 10$ MeV, $\Delta\mu_B = 10$ MeV, $D=0.06$ and $A = 0.5$ for 
    $s/n_B = 200,\, 100,\, 50,\, 33$ (from left to right). Dotted line represents the 
    estimated phase border line, long-dashed 
    lines in the left panel depict resonance gas trajectories with corresponding chemical 
    freeze-out points from~\cite{Becattini04}.}
  \vskip -.1cm
\end{figure}
seems to be generic (cf.~\cite{Nonaka05} for a 
different model with CEP which involves a hadronic low-temperature and a partonic 
high-temperature phase). 
Clearly, this statement decisively depends on $s$ in the hadronic phase where the simple toy model cannot 
account for QCD. 

In contrast, being inspired by lattice QCD we construct the EoS as truncated 
Taylor series expansion including the coefficients $c_{0,2,4,6}(T)$ of Fig.~1. 
These lattice based trajectories are shown in Fig.~\ref{fig:isentrops} (left panel) where the 
pattern differs notably from the observations made in the above toy model. Although first 
principle evaluations still suffer from too heavy quark masses and deviate consequently from the 
resonance gas trajectories in the low-temperature phase, the values $s/n_B$ on these 
lattice QCD deduced curves agree with the values at the according nearby chemical freeze-out 
points inferred from data~\cite{Becattini04}. 
Furthermore, in the deconfined phase lattice results are trustable due to convergence radius 
studies~\cite{All05} at least up to $\mu_B=300$ MeV. Therefore, adjusting the analytic 
contribution of the EoS known from the lattice data by our quasi-particle model, critical end point 
effects should become visible only for larger $\mu_B$ implying a small critical region. 
We include the CEP in line with the procedure outlined above in our quasi-particle model 
replacing $\bar{c}_2$ by $c_2(T)$ in $s_n$ and considering a small critical 
region characterized by $\Delta T = 10$ MeV, $\Delta \mu_B = 10$ MeV and $D = 0.06$ with strength 
parameter $A = 0.5$. As exhibited in Fig.~\ref{fig:isentrops} (right panel), CEP effects on isentropic 
trajectories are significant only for large $\mu_B$ with negligible impact on the hadronic sections. 
Nonetheless, the baryon number susceptibility $\chi_B = \partial^2 p / \partial\mu_B^2$ 
being a measure for baryon number fluctuations diverges for $\mu_B > \mu_{B,c}$ (Fig.~4 right panel) 
due to the discontinuity evolving in $n_B$ in contrast to the analytic behaviour (Fig.~4 left panel) 
stemming from the quasi-particle model not containing CEP effects. 
\begin{figure}[h]
  \vskip -.5cm
  \label{fig:suscept}
  \includegraphics[scale=0.28,angle=-90.]{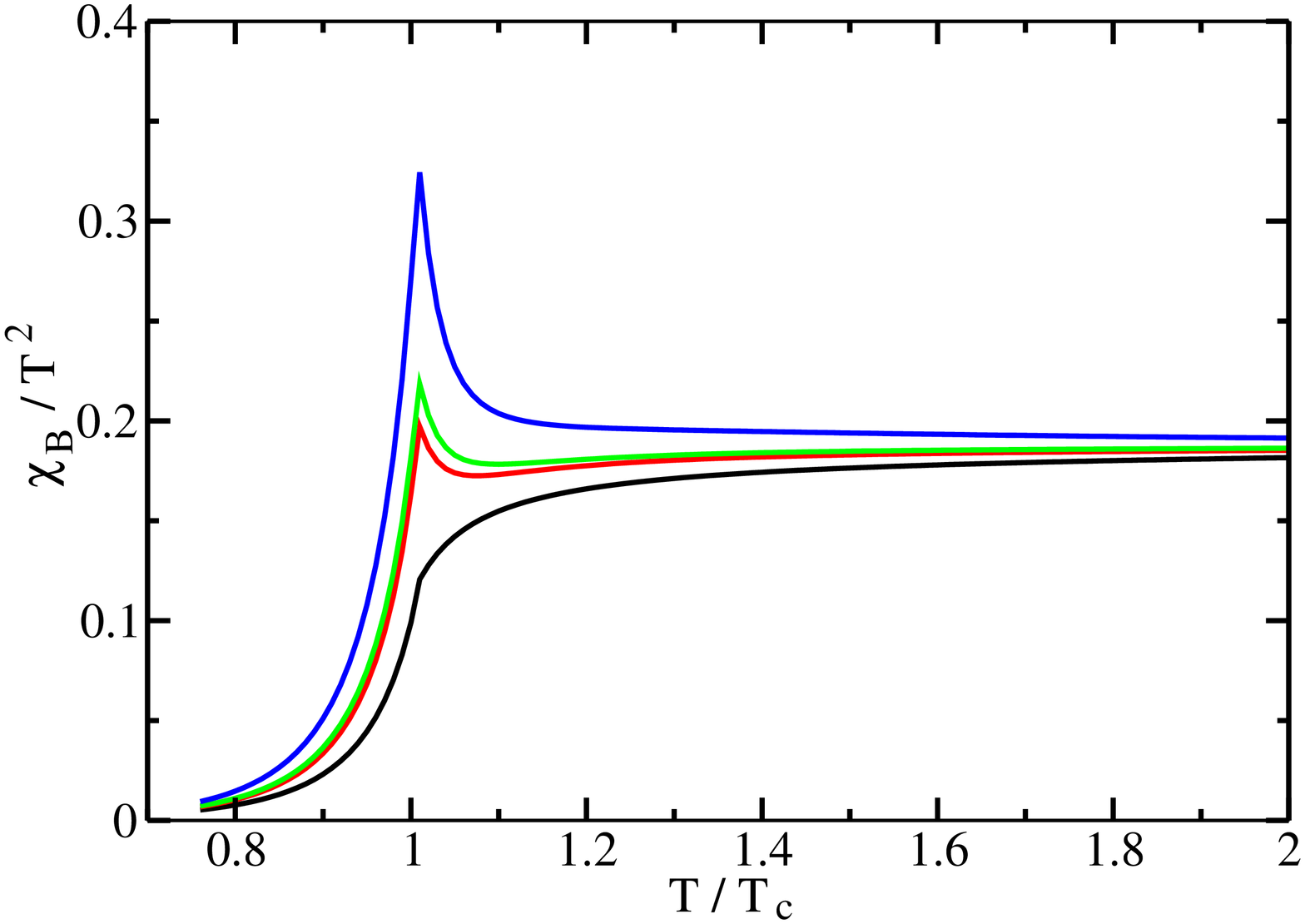}
  \includegraphics[scale=0.28,angle=-90.]{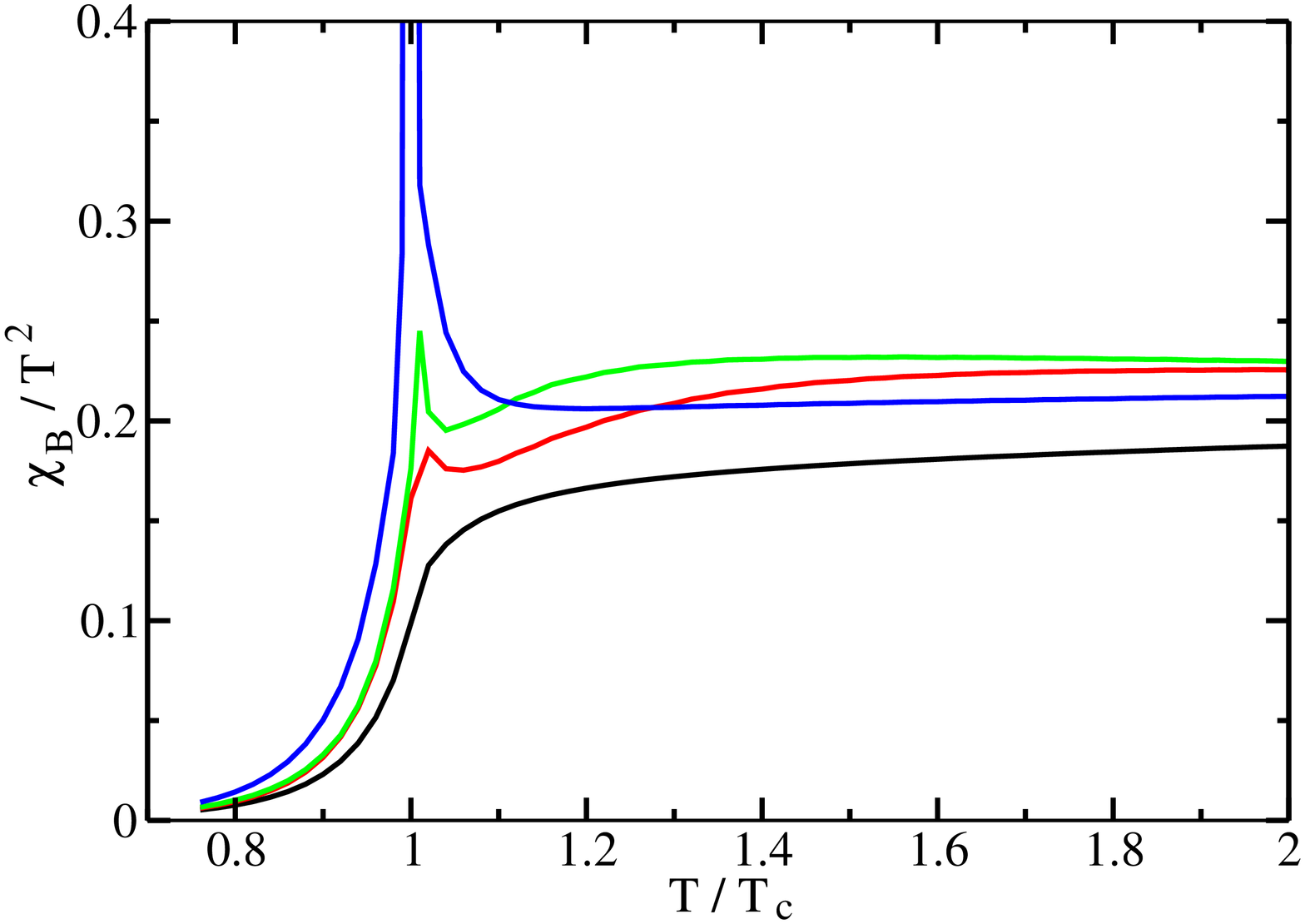}
  \caption{Scaled baryon number susceptibility of the quasi-particle model neglecting $c_6(T)$ 
    without CEP (left panel) and with CEP inclusion (right panel) as function of $T/T_c$ for 
    $\mu_B = 450,\, 330,\, 300,\, 150$ MeV (from top to bottom).}
\end{figure}

\section{Conclusion}
\label{sec:conclusions}

Our quasi-particle model without implemented CEP was successfully compared with 
recent lattice QCD results of the Taylor series expansion coefficients $c_0(T)$ and $c_{2,4,6}(T)$. 
Accordingly, our extrapolation procedure into the finite chemical potential region was tested. 
We considered simple models including phenomenologically the QCD critical end point and studied 
the effects on isentropic trajectories. We followed~\cite{Nonaka05} and looked for indications of the 
CEP acting generically as attractor or repulsor. In fact, this is of interest with respect to the 
question whether CEP effects show up in heavy ion experiments only in a very narrow beam energy 
range. Clearly, appropriate dynamical simulations are needed to account properly for such questions. 
The study of the pattern of isentropic trajectories is only a first step towards elucidating possible 
implications of the very existence of the CEP in QCD. 

Inspiring discussions with M.~Asakawa and F.~Karsch are greatfully acknowledged. The work is 
supported by BMBF, GSI, EU-I3HP. 

\end{document}